\documentclass[prl,aps,showpacs,superscriptaddress,twocolumn]{revtex4}
\usepackage{graphicx}
\usepackage{amsmath}
\usepackage{amsfonts}
\usepackage{amssymb}
\usepackage{epsfig}
\begin{document}

\newcommand\lavg{\left\langle}
\newcommand\ravg{\right\rangle}
\newcommand\ket[1]{\left|#1\right\rangle}
\newcommand\bra[1]{\left\langle#1\right|}
\newcommand\braket[2]{\left.\left\langle#1\right|#2\right\rangle}

\title{
Entanglement localization by a single defect in a spin chain}
\author{Tony J. G. Apollaro}\affiliation{
Dipartimento di Fisica, Universit\`a della Calabria, 87036
Arcavacata di Rende (CS) Italy}\affiliation{INFN - Gruppo
collegato di Cosenza, 87036 Arcavacata di Rende (CS) Italy}
\author{Francesco Plastina}\affiliation{ Dipartimento di Fisica, Universit\`a della
Calabria, 87036 Arcavacata di Rende (CS) Italy} \affiliation{INFN
- Gruppo collegato di Cosenza, 87036 Arcavacata di Rende (CS)
Italy}
\date{\today}

\begin{abstract}
We discuss the effect of a single diagonal defect on both the
static and dynamical properties of entanglement in a spin chain.
We show that entanglement localizes at the defect and discuss its
localization length, arguing that this can be used as a mean to
store entanglement. We also show that the impurity site can behave
as an entanglement mirror and characterize the bouncing process in
terms of reflection and transmission coefficients.
\end{abstract}
\bigskip
\pacs{03.67.Hk, 03.67.Mn, 05.50+q} \maketitle Entanglement
generation and distribution is a problem of central importance in
performing quantum-information tasks, like teleportation and
quantum cryptography. As entanglement between parties is created
by means of a direct interaction, methods are required to transfer
either the entangled particles or their state. Indeed, the problem
of designing quantum networks enabling efficient high-fidelity
transfer of quantum states has been addressed by a number of
authors, \cite{bose,superco,forcira,burbose,locmeas,chris},
especially focusing on the requirement of minimal control. This
means that state transfer should be possible without performing
many control operations such as switching on and off the
interactions, measuring, or applying encoding and decoding
procedures.

In this respect, spin systems provide ideal models to study
entanglement properties as they are naturally employed as qubit
registers and exploited as quantum channels (or coherent data
bus). Spin chains with fixed interactions have been
considered~\cite{bose}, and solid state implementations have been
already proposed \cite{superco}. It has also been shown that
perfect transfer can be achieved by performing local measurements
\cite{forcira}, by using several spin chains in parallel
\cite{burbose}, by employing local memories at the receiver
side~\cite{locmeas}, or by means of a spatial modulation of the
spin coupling strengths \cite{chris}. For this latter case, the
effect of imperfections (static errors) in the engineering of the
spin system has been analyzed in Ref. \cite{gabriele}, where a
scaling of the transmission fidelity has been obtained in terms of
the degree of imperfection. This is an example in which the
disorder has been taken into account. In fact, considerable
attention has been devoted to disordered spin systems. A threshold
in the coupling has been found for the onset of entanglement
between a bulk impurity and its neighborhood~\cite{wang}. Also it
has been shown that a faster (super-ballistic) distribution of
entanglement between a central node and a series of distributed
sites occurs if the sender resides in a disordered
region~\cite{fitzsi}. Furthermore, perturbative, numerical, and
Bethe-ansatz-based investigations of entanglement between two
defects has been performed for a disordered anti-ferromagnetic
spin-1/2 chain with anisotropic exchange coupling, \cite{lea}. A
possibility of tuning the ground state entanglement by a single
off-diagonal impurity in the anisotropic $XY$ model has also been
considered~\cite{osenda}.

Diagonal disorder has also been considered in the quantum
information context, but only the wave function localization issue
has been addressed in terms of the inverse participation
ratio~\cite{lfsantos}, while entanglement localization has been
completely overlooked, apart from some considerations reported in
Ref. \cite{santorigo}.

It is well known that disorder can lead to a spatial localization
of the electronic wave function~\cite{anderson} and that, in
particular, this occurs whenever impurities exist in a tight
binding model. In fact, a single impurity suffices \cite{economou}
and, in this paper, we discuss how this translates into an
entanglement localization.

We consider the simplest example of a system displaying
localization; namely, a spin chain with a single diagonal defect,
that is, a chain subject to a local static magnetic field which is
equal at every spin location except for a single site, where a
field defect is present. We analyze both the static and dynamical
properties of such a system, and, in particular, we discuss the
case in which the ground state of the chain is a localized state,
showing a very peculiar spatial distribution of quantum
correlations. We also discuss how this localized state affects the
possibility of sending entanglement through such a `defected'
channel. In particular, we show that a mirror-like effect occurs
at the defect during entanglement propagation. This achieves a
(partial) control of the entanglement dynamics, as, starting from
a homogeneous system, the modification of the single-qubit level
spacing permits the generation of an ``artificial defect".

The model we study is a 1-D isotropic XY spin-$\frac{1}{2}$ closed
chain, placed in an external magnetic field which is homogeneous
everywhere but for a single (defect) site $l$
\begin{eqnarray}
H=h\sum_{i=-\frac{N}{2}}^{\frac{N}{2}}{\sigma}_{z}^{i}
-J\sum_{i=-\frac{N}{2}}^{\frac{N}{2}}
({\sigma}_{x}^{i}{\sigma}_{x}^{i+1}
+{\sigma}_{y}^{i}{\sigma}_{y}^{i+1})+\epsilon\,{\sigma}_{z}^{l}
\label{hamiltonian}\end{eqnarray} where $h$ is the magnetic field,
acting on every site but for site $l$, where the field is
$h+\epsilon$. Also, $J$ is a ferromagnetic coupling constant,
while $\sigma_{\alpha}$ are the Pauli matrices. We will consider
the impurity term as a perturbation and take as a reference the
results obtained for the homogeneous XY Hamiltonian
\cite{plastina,subra}. Since $[H, \sum_i \sigma_z^i]=0$, the total
Hilbert space can be divided into invariant subspaces labelled by
the number of spins which point upwards. We will concentrate on
the $0$ and $1$-excitation subspaces and denote by $\ket n$ the
state with the upward spin at site $n$.

Thorough, we consider the case $2h> J$. This implies that the
unperturbed problem ($\epsilon=0$)  has the factorized ground
state $|0\rangle^{\otimes N+1}$, whose energy we re-scale to zero.
In the Hilbert space sector with one reversed spin, the
unperturbed energy basis is
\begin{equation}
|k\rangle=\frac{1}{\sqrt{N+1}}\sum_{n}\exp\left[\frac{2 \pi i k
n}{N+1}\right]|n\rangle
\end{equation}
with energy $E_{k}=2h-J\cos(\frac{2 \pi k}{N+1})$, and  where $k=
-\frac{N}{2}, - \frac{N}{2} +1 , \ldots , \frac{N}{2}$.

In this sector, the Hamiltonian (\ref{hamiltonian}) is equivalent
to a tight binding problem with one diagonal impurity. This can be
solved with the help of the Green's function
technique~\cite{economou}.

\noindent The unperturbed Green operator is
\begin{equation}
{G}_{0}(z)=\frac{1}{z-{H}_{0}}= \sum_{k}\frac{|k\rangle\langle
k|}{z-E_{k}}.\end{equation} In the thermodynamic limit
($N\rightarrow \infty$), the index $k$ becomes continuous and the
states $|k\rangle$ give rise to a continuous energy band, with
$E_k \in \left[2h-J,2h+J\right]=I_b$. In this continuum limit, the
matrix elements of $G_0$ between localized states can be written
\begin{eqnarray}
&& G_{0}(r,s;z)=\frac{(-x+\sqrt{x^2-1})^{|r-s|}}
{J\sqrt{x^2-1}}, \; z\neq I_b \\
&& G^{\pm}_{0}(r,s;z)=\frac{(-x\pm i\sqrt{1-x^{2}})^{|r-s|}}{\pm i
J\sqrt{1-x^{2}}}, \;  z\in I_b
\end{eqnarray}
where $x=\frac{z-2h}{J}$.

The Green operator for the full hamiltonian can be expressed in
terms of the unperturbed one, and, due to the simple form of the
interaction hamiltonian, the Dyson series can be re-summed giving
an exact expression \cite{economou}
\begin{equation}\label{fullgreen}
{G}(z)={G}_{0}(z)+ {G}_{0}(z)|l\rangle\frac{2\epsilon}{1-2\epsilon
G_{0}(l,l;z)}\langle l|{G}_{0}(z) .
\end{equation}
The knowledge of the Green function (\ref{fullgreen}) allows us to
obtain the spectrum of the hamiltonian (\ref{hamiltonian}). It
turns out that the energy band of the continuous spectrum is
unaffected, as for $E\in [2h-J,2h+J]$ the Green function has a cut
in the complex plane. The associated eigenstates are given by
$|\Psi(E)\rangle=\sum_{n}a_{n}(E)|n\rangle$, with
\begin{equation}
a_{n}(E)=\frac{1}{\sqrt{N+1}}\left( e^{i\theta n}+\frac{\alpha \,
e^{i |\theta| |n-l|}}{i|\sin\theta|-\alpha} \, e^{i\theta l}
\right),
\end{equation}
where $\alpha=\frac{2\epsilon}{J}$ and $\cos \theta =(2h -
E)/J$.\newline Besides a distortion of the states within the band,
the perturbation produces the appearance of a localized
eigenstate, whose energy is given by the pole of the Green
function (\ref{fullgreen}), determined by the equation $2\epsilon
G_{0}(l,l;E_{loc})=1$. This state lies above or below the band,
depending on wether $\alpha$ is greater than zero or not, and its
energy is $E_{loc}=2h\mp J\sqrt{1+\alpha^{2}}$. Explicitly, the
state is given by $\ket{\Psi_{loc}} = \sum_n b_n \, \ket n$, with
\begin{equation}
b_{n}=\begin{cases}
-\frac{\sqrt{|\alpha|}}{\left(1+\alpha^{2}\right)^\frac{1}{4}}
\exp[-\xi(\alpha)|n-l|] & \alpha<0 \\
(-1)^{|n-l|}\frac{\sqrt{|\alpha|}}{\left(1+\alpha^{2}\right)^{\frac{1}{4}}}
\exp[-\xi(\alpha)|n-l|]& \alpha>0 \end{cases}
\end{equation} where
$\xi(\alpha)=-\ln\left({\sqrt{1+\alpha^{2}}-|\alpha|}\right)$.
$\xi^{-1}$ is the spatial extension (or localization length) of
$\ket{\Psi_{loc}}$ around the impurity site $l$ ($\xi$ is obtained
in Ref. \cite{mav} for the $XXZ$ model). For
$\frac{2h}{J\sqrt{1+\alpha^{2}}}<1$ with $\alpha <0$ (that is, for
negative and large enough $\alpha$), $E_{loc}$ is lower than zero
and $\ket{\Psi_{loc}}$ becomes the ground state of the system.

We analyze the entanglement content of this localized state. In
particular, we concentrate on the pairwise quantum correlations
shared by the qubits at sites $i$ and $j$ as measured by the
concurrence~\cite{wootters}. It is easy to show that the
concurrence between these sites is given by the residue of the
associated matrix element of the Green operator (to simplify the
notation, we set $l=0$ and measure distances along the chain from
the impurity site)
\begin{equation}
C_{ij}= 2 Res[G(i,j;E_{loc})]=\frac{2
|\alpha|}{\sqrt{1+\alpha^{2}}} \; e^{-\xi \, (|i|+|j|) }.
\label{cigei}
\end{equation}
As a result of the presence of the magnetic field defect, pairwise
entanglement is only present for spins residing within a distance
$1/\xi$ from the impurity and it decays exponentially with the
separation. In this sense, we state that entanglement is localized
in the neighborhood of the defect. As the localization length
becomes less than the site spacing for $\alpha \leq (1-e^2)/2
e\simeq -1.175$, entanglement is highly localized even for small
values of the defect field.

\begin{figure}
\includegraphics[width=0.5\textwidth]{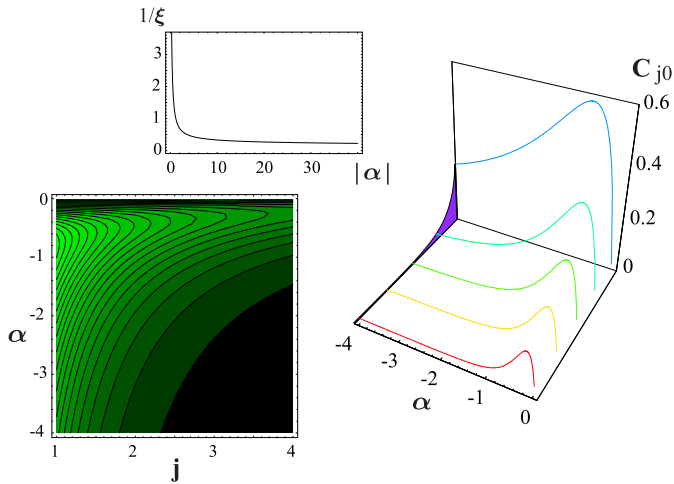}\\
\caption{(up-left) Localization length as a function of
$|\alpha|$. (down-left and right) Ground state concurrence between
the impurity and the spin at site $j$. Notice the exponential
decrease as a function of the distance. }\label{length}
\end{figure}
We now analyze the transmission of a quantum state along the
chain. In particular, we are interested in characterizing
entanglement distribution; namely, the possibility of using the
chain to send one partner of a maximally entangled pair. A related
problem is the transmission of a single qubit state along the
chain. To send one qubit, we imagine that at site $s$, the sender
prepares a generic single qubit state $\ket{\psi_s}$ which is
aimed to be retrieved at site $r$ (where, in general, the mixed
state $\rho_r(t)$ is extracted at time $t$). A measure of the
quality of the chain as a transmission channel is given by
fidelity, $F (t) = \bra{\psi_s} \rho_r (t) \ket{\psi_s}$, averaged
over all possible messages $\ket{\psi_s}$. On the other hand, to
send entanglement, the spin $s$ is initially prepared in a singlet
state with an external (un-coupled) qubit. The entanglement
transmission can be characterized by the (final) concurrence
between the external qubit and the one residing at the receiving
site $r$, denoted by $C_{r}(t)$.

In fact, the two quantities $F$ and $C$ are related \cite{bose},
as both can be expressed in terms of the transmission amplitude to
send an excitation from $s$ to $r$,  $f_{rs}(t) = \bra r e^{-i H
t} \ket s$. Explicitly, the averaged fidelity is $\langle F
\rangle(t)= (|f_{rs}(t)+1|^{2}+2)/6,$ whereas the concurrence is
given by $C_r (t) = |f_{rs}|$. In our case, the transition
amplitude is given by a complicated combination of matrix elements
of the Green functions,
\begin{eqnarray}\label{frs}
&& f_{rs}(t)=  \int_{-\pi}^{\pi}\frac{d\theta}{2\pi}\Biggl \{
e^{i\theta(r-s)}+e^{i\theta(l-s)} g^{(+)}_{r,l} +e^{i\theta(r-l)}
\, g^{(-)}_{l,s} \nonumber
\\ && \quad + g^{(+)}_{r,l} g^{(-)}_{l,s} \Biggr \} \, e^{-i E
t}+ Res[G(r,s;E_{loc})] \, e^{-i E_{loc}t}
\end{eqnarray}
where, within the integral, $E=2h - J \cos \theta$, while
$$g^{(\pm)}_{i,j} (E)= \frac{\alpha G_{0}^{(\pm)}(i,j;E)}{1-\alpha
G_{0}^{(\pm)}(l,l;E)} =\frac{\alpha\, e^{\pm i|\theta| |i-j|}}{\pm
i\sin{ |\theta|} -\alpha}.$$ The amplitude $f_{rs}$ given in eq.
(\ref{frs}) contains two terms. One contribution comes from the
localized state; it has the same structure of the concurrence
$C_{rs}$ given in eq. (\ref{cigei}). The other contribution is
written in terms of an integral over the branch cut region of the
Green function, describing the evolution of the states within the
continuous band. While the former goes to zero with the amplitude
of the defect $\alpha$, the latter does not, as it reduces to the
un-perturbed transition amplitude when $\alpha \rightarrow 0$.
This is given by the Bessel function of order $r-s$ and argument
$\tau = J t$,$J_{r-s}(\tau)$, see Ref. \cite{plastina}. The main
effect of a finite defect field in the second term is a distortion
of the amplitude, with the appearance of contributions describing
scattering at the defect during propagation from $s$ to $r$.

In the limit of $\alpha \gg 1$, a series of simple remarks can be
drawn. First of all, the localization length goes to zero, and
therefore the localized state is really concentrated on the
impurity site. As a result, if the qubit residing there is
initially entangled with an external one, the entanglement doesn't
propagate at all, and the concurrence has the simple form (for
$s=l=0$):
\begin{equation}
C_r \simeq \begin{cases} 1 - \frac{1}{2 \alpha^2} & r = 0 \\
 \frac{r}{|\alpha| \, \tau} \, J_r(\tau) & r\ne 0 \end{cases}
\end{equation}
Secondly, the transmission through the defect is very poor.
Indeed, when the sender is on one side of the impurity (located at
$l=0$), while the receiving site is on the other one, at a
distance $d=|r|+|s|$, the concurrence is given by
\begin{equation}C_r \simeq \frac{1}{2 |\alpha|} \Bigl
[J_{d+1}(\tau) + J_{d-1}(\tau) \Bigr ],\end{equation} which decays
both with the distance between the sites and with time as
$\tau^{-3/2}$.

Finally, the defect can reflect entanglement. Indeed, when both
the sender and the receiver are on the same side (say $s,r>0$ with
$r>s$), the concurrence is given by the combination of three
terms:
\begin{equation}
C_r= \left |(-1)^s J_{r-s}(\tau) - J_{r+s} - i \frac{J_{r+s+1}+
J_{r+s-1}}{2 \alpha} \right |.\end{equation} This relation has a
simple interpretation: the entanglement can either propagate
directly from $s$ to $r$ (a process weighted by the un-perturbed
transition amplitude given by the Bessel function $J_{r-s}$), or
it can propagate to the defect and then be reflected to the site
$r$ (thus travelling a distance $2s + r-s = r+s$). This latter
process has an amplitude obtained by {\it i)} taking the Bessel
function $J_{r+s}$ which describes propagation along this
distance, and {\it 2)} subtracting (with its own relative phase)
the amplitude for entanglement transmission through the defect.

\begin{figure}
\includegraphics[width=0.22\textwidth]{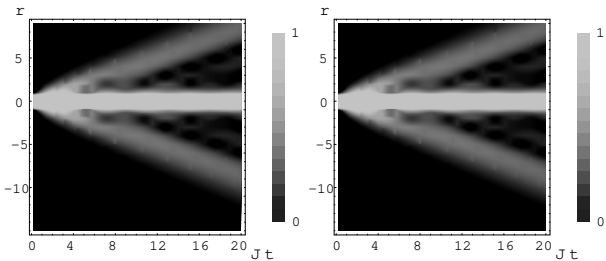}
\includegraphics[width=0.22\textwidth]{locaar.eps}\\
\caption{Time evolution of the concurrence $C_r$ for $\alpha=-2$
and two different initial conditions. The left plot shows the case
in which the sender resides at the defect site ($l=s=0$), whereas
the right plot shows the case $s=-5$. In the first case, the
entanglement with the external qubit stays localized at the sender
site, while in the second one, entanglement propagates almost
symmetrically from the sender site up to the defect, where it is
reflected back.}\label{loca}
\end{figure}
Although not easily seen from the full expression of eq.
(\ref{frs}), these two main features of entanglement localization
and reflection at the impurity site, are present also for moderate
values of $\alpha$. This can be seen in Fig. (\ref{loca}), where
the case $\alpha = -2$ is illustrated. By increasing $\alpha$,
these effects become more and more pronounced and the secondary
propagation lines which are present in Fig. (\ref{loca}) tend to
disappear.
\begin{figure}
\includegraphics[width=0.35\textwidth]{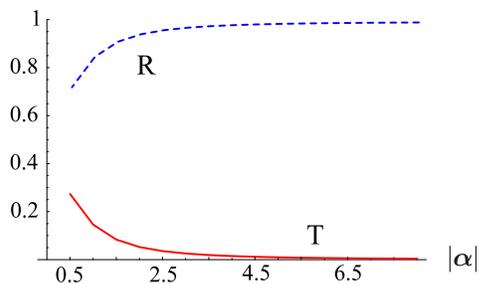}\\
\caption{Transmission and reflection coefficient as defined by Eq.
(\ref{erreti}), displayed as a function of
$\alpha$.}\label{refletr}
\end{figure}

To better characterize the effect of the impurity on entanglement
propagation, we can introduce reflection and transmission
coefficients. Indeed, we can exploit the fact that states with one
reversed spin saturate the CKW inequality \cite{ckw}, which in our
case implies that $\sum_r C_r^2 (t)=1$, and use the tangle as a
probability distribution to define (to be specific, we consider
the case of $s<0$)
\begin{equation}
T=\lim_{t\rightarrow \infty} \sum_{r>0} C_r^2(t), \quad
R=\lim_{t\rightarrow \infty} \sum_{r<0} C_r^2(t). \label{erreti}
\end{equation}
The transmission and reflection coefficients are shown in Fig.
(\ref{refletr}) as functions of $\alpha$.

Entanglement reflection at the defect is completely accounted for
by the presence of the localized state. Indeed, since the overlap
between the initial state and $\ket{\Psi_{loc}}$ is exponentially
small, entanglement propagates almost freely up to the defect
site, but then the spin-reversal excitation cannot enter there. As
the Hamiltonian describes nearest neighbor interaction, subsequent
spins cannot receive the excitation and are, therefore, almost
excluded from the dynamics. As a result, entanglement is reflected
backwards.

In summary, we discussed the effect of a single magnetic field
defect on entanglement in a spin chain. We found that this
impurity produces a state with localized entanglement, whose
presence strongly affects the dynamics of the chain. Indeed, the
qubit at the defect site can either maintain entanglement for long
time or work as an entanglement mirror. These two properties,
obtained by modifying the local magnetic field of just one qubit,
can be exploited to achieve a control on the propagation of
pairwise quantum correlations in distributed qubit systems.
Indeed, we have shown that entanglement can be stored for long
times at the defect site and that its propagation can be modified
so that it can be directed towards the receiver.

This work has been done within the framework of the PRIN2005029421
project.


\begin{thebibliography}{99}
\bibitem{bose}
S. Bose, Phys. Rev. Lett. {\bf 91}, 207901 (2003).
%
\bibitem{superco}
A. Romito, R. Fazio, and C. Bruder, Phys. Rev. B {\bf 71}, 100501
(2005); M. Paternostro, G. M. Palma, M. S. Kim, and G. Falci,
Phys. Rev. A {\bf 71}, 042311 (2005).
%
\bibitem{forcira}
F. Verstraete, M.A. Mart\`{\i}n-Delgado, and J. I. Cirac, Phys.
Rev. Lett. {\bf 92} 087201 (2004).
%
\bibitem{burbose}
D. Burgarth, and S. Bose, Phys. Rev. A {\bf 71}, 052315 (2005).
%
\bibitem{locmeas}
V. Giovannetti and D. Burgarth, Phys. Rev. Lett. {\bf 96}, 030501
(2006).
%
\bibitem{chris}
M. Christandl, N. Datta, A. Ekert, and A. J. Landahl, Phys. Rev.
Lett. {\bf 92}, 187902 (2004); M. Christandl, N. Datta, T. C.
Dorlas, A. Ekert, A. Kay, and A. J. Landahl, Phys. Rev. A {\bf
71}, 032312 (2005); P. Karbach and J. Stolze, Phys. Rev. A {\bf
72}, 030301(R) (2005).
%
\bibitem{gabriele}
G. De Chiara, D. Rossini, S. Montangero, and R. Fazio, Phys. Rev.
A {\bf 72}, 012323 (2005).
%
\bibitem{wang}
X. Wang, Phys. Rev. E {\bf 69}, 066118 (2004).
%
\bibitem{fitzsi} J.
J. Fitzsimons, and J. Twamley, Phys. Rev. A {\bf 72}, 050301
(2005).
%
\bibitem{lea}
L.F. Santos, Phys. Rev. A {\bf 67}, 062306 (2003); L.F. Santos,
and G. Rigolin, Phys. Rev. A {\bf 71}, 032321 (2005).
%
\bibitem{osenda}
O. Osenda, Z. Huang, and S. Kais, Phys. Rev. A {\bf 67}, 062321
(2003).
%
\bibitem{lfsantos}L.F. Santos, M.I. Dykman, M. Shapiro, and F.M. Izrailev, Phys.
Rev. A {\bf 71}, 012317 (2005).
%
\bibitem{santorigo}
L.F. Santos, G. Rigolin, and C.O. Escobar, Phys. Rev A {\bf 69},
042304 (2004).
%
\bibitem{anderson}
P. W. Anderson, Phys. Rev. {\bf 109}, 1492 (1958).
%
\bibitem{economou}
E. N. Economou, "Green's Function in Quantum Physics" 2-nd
edition, Springer-Verlag, Berlin (1983).
%
\bibitem{plastina} L. Amico, A. Osterloh, F. Plastina, R.
Fazio, and G. M. Palma, Phys. Rev. A, {\bf 69}, 022304 (2004).
%
\bibitem{subra}
V. Subrahmanyam, Phys. Rev. A, {\bf 69}, 034304 (2004).
%
\bibitem{mav}
L. F. Santos, and M. I. Dykman, Phys Rev. B {\bf 68}, 214410
(2003).
%
\bibitem{wootters}
W. K. Wootters, Phys. Rev. Lett. {\bf 80}, 2245
(1998).
%
\bibitem{ckw} V. Coffman, J. Kundu, and W. K. Wootters,
Phys. Rev. A {\bf 61}, 052306 (2001).

\end{thebibliography}
\end{document}